\documentclass{article}
\usepackage{spconf,amsmath,graphicx}
\usepackage{color}
\usepackage{multirow}
\usepackage{bm}
\usepackage{footnote}
\usepackage{algorithm}  
\usepackage{algorithmicx}  
\usepackage{algpseudocode}  
\usepackage{hyperref}
\RequirePackage{booktabs}

\title{Time Domain Audio Visual Speech Separation}
\name{Jian Wu$^{1,2^*}$, Yong Xu$^3$, Shi-Xiong Zhang$^3$, Lian-Wu Chen$^2$, Meng Yu$^3$, Lei Xie$^1$, Dong Yu$^3$}

\address{
  $^1$School of Computer Science, Northwestern Polytechnical University, Xi'an, China\\
  $^2$Tencent AI Lab, Shenzhen, China \thanks{$^*$ This work was done when Jian Wu was an intern in Tencent AI lab.} \\
  $^3$Tencent AI Lab, Bellevue, USA}

\begin{document}
%
\maketitle

\begin{abstract}
Audio-visual multi-modal modeling has been demonstrated to be effective in many speech related tasks, such as speech recognition and speech enhancement. This paper introduces a new time-domain audio-visual architecture for target speaker extraction from monaural mixtures. The architecture generalizes the previous TasNet (time-domain speech separation network) to enable multi-modal learning and at meanwhile it extends the classical audio-visual speech separation from frequency-domain to time-domain. The main components of proposed architecture include an audio encoder, a video encoder that extracts lip embedding from video streams, a multi-modal separation network and an audio decoder. Experiments on simulated mixtures based on recently released LRS2 dataset show that our method can bring 3dB+ and 4dB+ Si-SNR improvements on two- and three-speaker cases respectively, compared to audio-only TasNet and frequency-domain audio-visual networks.
\end{abstract}

\section{Introduction}
\label{intro}

The goal of speech separation is
to separate each source speaker from the mixture signal. Although it has been studied for many years, speech separation is still a difficult problem, especially in noisy and reverberated environment. Several audio-only speech separation methods were recently proposed, such as uPIT \cite{kolbaek2017multitalker}, DPCL \cite{hershey2016deep, isik2016single}, DANet \cite{chen2017deep} and TasNet \cite{luo2018tasnet}. 
However, in these approaches, the number of target speakers has to be known as a prior information and assumed to be unchanged during training and testing. In addition, the separation results of these systems cannot be associated to the speakers, which greatly limits their application scenarios.


If we can extract some target speaker dependent features, the task of speech separation will become the target speaker extraction problem. This has several clear advantages over the blind separation approaches. 
First, as the model only extracts one target speaker each time from the mixture, the prior knowledge about the number of speakers is no longer needed. Second, as the target speaker features were given to the model, the issue of label permutation is apparently avoided. These make separation of the target speaker more practical than the blind separation solutions.

Several target speaker separation approaches have been explored in the past \cite{zmolikova2017speaker, wang2018deep, wang2018voicefilter}. In \cite{wang2018voicefilter}, the authors proposed a system named VoiceFilter that used d-vectors \cite{wan2018generalized} as the embedding of target speaker for the separation network. Similarly \cite{wang2018deep} used a short anchor utterance as auxiliary input for target speaker separation. However, speaker or utterance based features are not robust enough, which could heavily affect system performance, especially when noise exists or same-gender speakers are mixed.

Alternatively, 
the visual information is acoustic noise insensitive and highly correlated to the speech content. 
Combining the audio and visual information has previously been investigated for automatic speech recognition (ASR) \cite{chung2017lip, afouras2018deep, shillingford2018large}, speech enhancement \cite{gabbay2018visual, gabbay2018seeing} and speech separation \cite{afouras2018conversation, ephrat2018looking}. The results have shown great potential of making use of visual features as complementary information source. For speech separation, \cite{gabbay2018visual} proposed an encoder-decoder model architecture to separate the voice of a visible speaker from background noise. The noisy spectrogram and center cropped video frames were used as the input of the audio/visual encoder respectively. \cite{afouras2018conversation} built a deeper network via stacking depth-wise separable convolution blocks, which includes a magnitude network to estimate Time-Frequency (TF) masks \cite{wang2018supervised} and a phase network to predict clean phase from the mixture phase. The magnitude network used pre-trained lip embeddings \cite{stafylakis2017combining} as visual features. 
Similarly \cite{ephrat2018looking} built a model based on dilated convolutions and LSTMs using face embeddings as visual features and yielded good results on a large scale audio-visual dataset.

Most of previous audio-visual separation systems have handled the audio stream in TF domain and thus the accuracy of the estimated TF-mask is the key to the success of those systems. On the other hand, the phase of signals is less considered, although it can significantly affect separation quality \cite{wang2018end, williamson2016complex,le2019phasebook,wang2019deep}. To incorporate the phase information, \cite{afouras2018conversation} used a phase subnet to refine original noisy phase, and \cite{ephrat2018looking} adopted newly proposed complex masks \cite{williamson2016complex} instead of traditional real masks (e.g., IRM \cite{kolbaek2017multitalker}, PSM \cite{erdogan2015phase} etc). 
Recently, \cite{luo2018tasnet, luo2019conv} proposed a new encoder-decoder framework named TasNet, directly separating speech on time-domain with impressive results on public WSJ0-2mix dataset. In this paper, we propose a new separation network 
that generalizes the TasNet to enable multi-modal fusion of auditory and visual signals. 
Given a raw waveform of mixture speech and the corresponding video stream of the target speaker, our audio-visual speech separation model can extract audio of the the target speaker directly.

The contribution of this work includes: 1) A new structure for multi-modal speech separation is proposed and to illustrate the effectiveness of the structure, a comprehensive comparison \footnote{Audio samples of the compared systems can be checked out at \url{https://funcwj.github.io/online-demo/page/tavs}} is performed with three typical separation models, uPIT \cite{kolbaek2017multitalker} (frequency-domain audio-only), Conv-TasNet \cite{luo2018tasnet} (time-domain audio-only) and Conv-FavsNet (frequency-domain audio-visual). 2) To the best of our knowledge, this is the first work that performs audio-visual separation directly on the time-domain. Experiments on recently released in-the-wild videos \cite{afouras2018deep} show that the proposed structure brings significant improvements compared to all other baseline models. 3) Previous visual features are not well designed for speech separation. In this work, the visual (lip) embeddings are specifically trained to represent the phonetic information. Different modeling units for the embedding network, such as words, phonemes (CI-phone) and context-dependent phones (CD-phone), are also investigated and compared.



\section{Proposed system}

\begin{figure*}[!tbp]
\centering
\includegraphics[width=0.9 \textwidth]{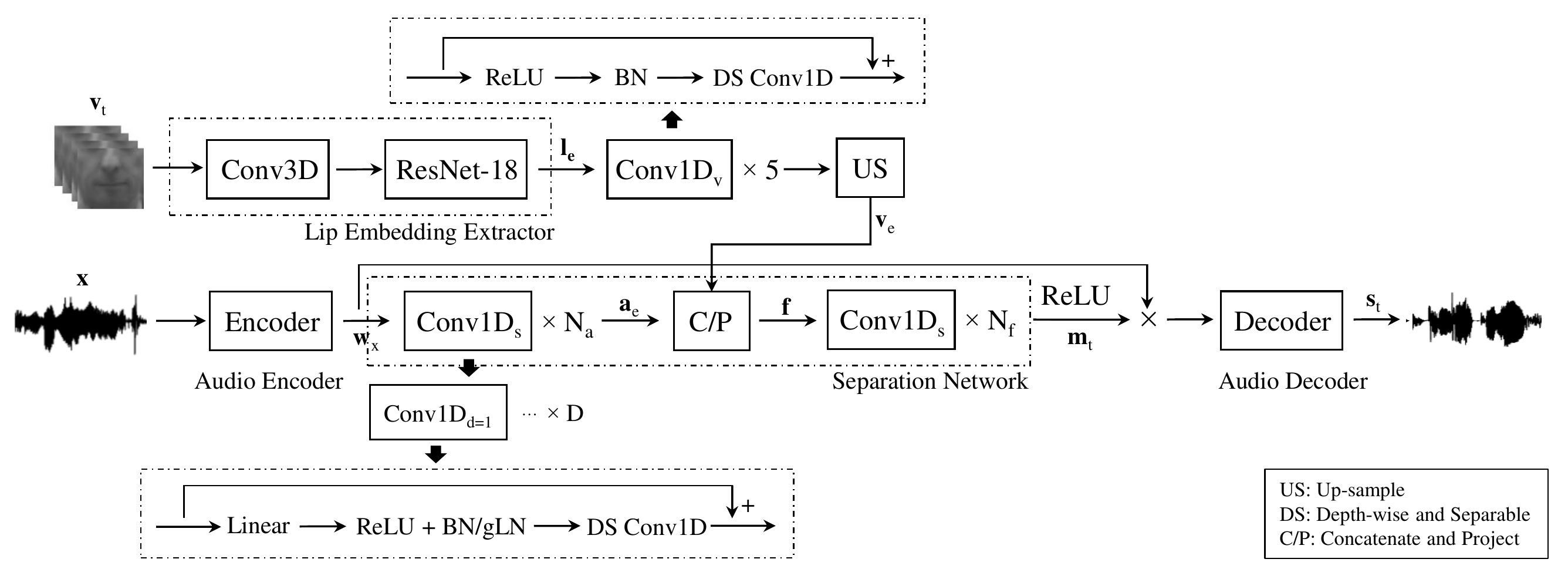}
\caption{Proposed time-domain audio-visual separation network. The lip region of each input image frame in $\mathbf{v}_t$ is cropped out and resized to 112 $\times$ 112 before feeding into the spatiotemporal convolutional block $\text{Conv3D}$ \cite{stafylakis2017combining}. Lip embedding $\mathbf{l}_e$, audio encoder output $\mathbf{w}_x$, video encoder output $\mathbf{v}_e$, audio encoded feature $\mathbf{a}_e$ and fused feature $\mathbf{f}$ are 256 dimensional sequences.}
\label{model}
\end{figure*}

\label{sec2_description}
This section will introduce the architecture of our proposed time-domain audio-visual speech separation network. Generally speaking, the network is fed with chunks of raw waveform and corresponding video frames, and predicts the speech of target speaker directly, as depicted in Figure \ref{model}. Scale-invariant source-to-noise ratio (Si-SNR) is used as the training objective function \cite{luo2018tasnet}, which is defined as
\begin{equation} \label{}
    \text{Si-SNR} = 20 \log_{10} \frac{\Vert \alpha \cdot \mathbf{s}_t \Vert}{\Vert \mathbf{s}_e - \alpha \cdot \mathbf{s}_t \Vert}.
\end{equation}
$\mathbf{s}_e, \mathbf{s}_e$ are estimated signal and target source respectively and are normalized to zero mean, where $\alpha$ is an optimal scaling factor computed via
\begin{equation} \label{}
\alpha = \mathbf{s}_e^T\mathbf{s}_t / \mathbf{s}_t^T\mathbf{s}_t.
\end{equation}

We also use Si-SNR as the evaluation metric, which is thought to be a more robust measure of separation quality compared to the original SDR \cite{le2019sdr}.

\subsection{Overview}
The proposed structure is mainly inspired by the TasNet structure proposed in \cite{luo2019conv}, which contains three parts, an audio encoder/decoder and a separation network. Given an audio mixture chunk $\mathbf{x} = \{x_0, x_1, \cdots, x_C\}$ ($C$ means the chunk size), the audio encoder in the TasNet tries to encode mixture samples as some non-negative vector sequences $\mathbf{w}_x$, while the separation network estimates masks of each source defined on such space. The audio decoder is used to reconstruct each masked results into time-domain again. The total framework could be described as
\begin{equation} \label{}
    \mathbf{s}_i = \text{Decoder}(\mathbf{w}_x \odot \mathbf{m}_i),
\end{equation}
where 
\begin{equation} \label{}
\begin{aligned}
    \mathbf{w}_x &= \text{Encoder}^a(\mathbf{x}), \\
    [\mathbf{m}_0, \cdots, \mathbf{m}_{N-1}] & = \text{Separator}(\mathbf{w}_x).
\end{aligned}
\end{equation}
$\odot$ denotes the Hadamard product. $N$ is the number of the speaker sources in the mixture. $\mathbf{m}_i$ and $\mathbf{s}_i$ represent the estimated masks and separation results of the $i$-th speaker, respectively.

When the video stream is available, our motivation is to bias the separation network through feeding both audio representations $\mathbf{w}_x$ and video encoded features $\mathbf{v}_e$. Thus the separator only generates masks $\mathbf{m}_t$ of the target speaker, which could be written as
\begin{equation} \label{}
\begin{aligned}
\mathbf{v}_e &= \text{Encoder}^{v}(\mathbf{v}_t), \\
\mathbf{m}_t &= \text{Separator}(\mathbf{w}_x, \mathbf{v}_e).
\end{aligned}
\end{equation}
$\mathbf{v}_t$ means the image frame sequences of the target speaker, and $\text{Encoder}^{a/v}$ represent the audio/video encoder. Finally, separated results $\mathbf{s}_t$ could be obtained through the formula below:
\begin{equation} \label{}
\mathbf{s}_t = \text{Decoder}( \mathbf{w}_x \odot \mathbf{m}_t).
\end{equation}

\subsection{Video encoder}

Video encoder is designed to extract the visual features $\mathbf{v}_e$ from input image frame sequences. In our experiments, we only use the lip region because it encodes the context and phonetic information we need. Generally, the video encoder contains a lip embedding extractor, followed by several temporal convolutional blocks, as depicted in Figure \ref{model} .

The lip embedding extractor consists of a 3D convolution layer and a 18-layer ResNet \cite{he2016deep}, similar to the work in \cite{afouras2018conversation}. The extractor outputs the fixed dimensional feature vectors $\mathbf{l}_e$ for each video frame and $\mathbf{l}_e$ is passed through the following temporal convolutional blocks. Each block consists of a temporal convolution, preceded by ReLU activation and batch normalization \cite{ioffe2015batch}. The residual connection is also included, although they do not have significant impact according to our results. To reduce the model parameters, we use depth-wise separable convolution \cite{chollet2017xception} instead. In our setups, we use a 256 dimensional lip embeddings and choose 3 kernel size, 1 stride size and 512 channel size for all the blocks.

The lip embedding extractor is pre-trained separately with specific back-end, following the steps introduced in \cite{stafylakis2017combining}. After that, the extractor is kept frozen. Apart from the word-level classification target used in \cite{afouras2018conversation}, we also try the phoneme (CI-phone) and context dependent phone (CD-phone) from a view of ASR to improve the quality of the lip embeddings.



\subsection{Audio encoder/decoder}
Audio encoder and decoder perform the 1D convolution and deconvolution operation on the mixed audio signals $\mathbf{x}$ and the masked encoded sequences, respectively. They can be represented as:
\begin{equation} \label{}
\begin{aligned}
 \text{Encoder}^a(\mathbf{x})  & = \text{ReLU}(\text{conv}_{\text{1D}}(\mathbf{x}, K, S)) \\
 \text{Decoder}(\mathbf{x}) &= \text{deconv}_{\text{1D}}(\mathbf{x}, K, S)
\end{aligned}
\end{equation}
$K$ and $S$ denote size of kernel and stride in 1D convolution operation, respectively. In this paper, we use $K=40$ and $S=20$ by default.

\subsection{Separation network}
Separation network is designed for estimating masks $\mathbf{m}_t$ of the target speaker, conditioned on the encoded audio and visual features. It is stacked by several temporal dilated convolutional blocks and provides a simple mechanism to handle feature fusion and synchronization. The structure of the 1D dilated convolutional block in separation network is similar to the one used in Conv-TasNet. In this paper, we denote it as $\text{Conv}_s$. Each $\text{Conv}_s$ has D sub-blocks with the exponential growth dilation factors $2^d$, where $d \in \{0, \dots, D - 1\}$, as shown in Figure \ref{model}. In our setups, we use $D = 8$.

The output of the audio encoder $\mathbf{w}_x$ are firstly passed through $N_a$ convolutional blocks
\begin{equation} \label{}
    \mathbf{a}_e = \overbrace{\text{Conv}_s(\cdots\text{Conv}_s(\mathbf{w}_x))}^{N_a}
\end{equation}
and then fused with visual features $\mathbf{v}_e$. The fusion process is performed through a simple concatenation operation over the convolution channel dimensions, followed by a position-wise projection $\mathcal{P}$ to reduce the feature dimension. In order to synchronize the time resolution of audio and video features, up-sampling is done on video streams before concatenation if it's necessary. The description above could be written as:
\begin{equation} \label{}
    \mathbf{f} = \mathcal{P}([\mathbf{a}_e; \text{Upsample}(\mathbf{v}_e)])
\end{equation}
Finally, the fused features $\mathbf{f}$ are fed through $N_f$ convolutional blocks and target masks $\mathbf{m}_t$ are estimated:
\begin{equation} \label{}
    \mathbf{m}_t = \sigma(\overbrace{\text{Conv}_s(\cdots\text{Conv}_s(\mathbf{f}))}^{N_f}).
\end{equation}
$\sigma$ here means an arbitrary non-linear function. We use $\text{ReLU}$ in our experiments, without loss of generality.

We mainly utilize two normalization techniques in the separation network, batch normalization (BN) and global layer normalization (gLN). gLN normalizes the features on both time and channel dimensions. Although the modules applied gLN can not be a causal system, it brings better performance in the practice of the Conv-TasNet.

\section{Experiments and results}
\label{sec3_exp}

\begin{table}
\centering
\caption{Number of utterances used to simulate each dataset for 2 and 3-speaker mixtures}
\label{tab1}
\begin{tabular}{cccc}
\toprule
Dataset     & \#utts     & Source          & \#simu  \\ \midrule
training    & 16445     & \emph{training}   & 40k    \\ 
validation  & 3000      & \emph{training}   & 5k     \\ 
test        & 740       & \emph{validation}+\emph{test} & 3k \\ 
\bottomrule
\end{tabular}
\end{table}

\subsection{Dataset}
In the experiment, we created two-speaker and three-speaker mixtures using utterances from Oxford-BBC Lip Reading Sentences 2 (LRS2) dataset~\cite{afouras2018deep}, which consists of thousands of spoken sentences from BBC television with their corresponding transcriptions. The \emph{training}, \emph{validation} and \emph{test} sets are generated according to the broadcast date, and thus those sets are not overlapped.

Short utterances (less than 2s) are dropped and 21075 utterances in total are used for data generation, with 19445 for training and the rest for validation and testing, respectively. The details of utterances used for simulation are summarized in Table \ref{tab1}. Two- and three-speaker mixtures are generated by randomly selecting different utterances and mixing them at various signal-to-noise ratios (SNR) between -5 dB and 5 dB. The sampling rate is 16kHz. To ensure the videos of each source are available in a mixture, longer sources are truncated to be aligned with the shortest one. The source segment is synchronous with the video stream in 25 \emph{fps}. Finally, we simulated 40k (25h+ in total), 5k and 3k utterances for training, validation and test set, respectively. 

\begin{savenotes}
\begin{table}
\centering
\caption{Results of oracle TF-mask and audio-only methods on 2 and 3-speaker test set}
\label{tab2}
\begin{tabular}{ccccc}
\toprule
	\multirow{2}{*}{Method} & \multirow{2}{*}{Configuration} & \multirow{2}{*}{\#Param} & \multicolumn{2}{c}{Si-SNR}       \\
	             &  &  &  2spkr & 3spkr      \\  \midrule
	 Mixed & - & - & 0.01 & -3.33 \\
	Oracle PSM        & 512/256/hann  & - & 14.43   &  11.54   \\ 
	Oracle IRM        & 512/256/hann  & - &  11.33  &  8.18   \\ 
	uPIT-BLSTM\footnote{\url{https://github.com/funcwj/setk/tree/master/egs/upit}}       & 3$\times$600  & $\sim$22M & 7.13 & 1.20 \\
	Conv-TasNet\footnote{\url{https://github.com/funcwj/conv-tas-net}} & gLN & $\sim$13M & 10.58 &  5.67   \\ 
\bottomrule
\end{tabular}
\end{table}
\end{savenotes}

\subsection{Training details}
Similar to \cite{afouras2018conversation}, we first train the lip embedding extractor on the LRW dataset. This is a word-level classification task and we achieve 76.02\% classification accuracy on the test set \cite{stafylakis2017combining}. However, the LRW dataset does not provide utterance-level audio transcripts, which makes it unfeasible to replace the word-level training targets with smaller pieces, e.g., CD/CI-phones mentioned above. Instead, we choose LRS2's \emph{pre-train} set to train the phone-level lip embedding extractor. 

The alignments are derived from the GMM acoustic models following the Kaldi's \cite{povey2011kaldi} recipes and are sub-sampled to the video sampling rate before training. We choose 44 units from CMU dictionary for CI-phones and get 3048 units from alignments set for CD-phones. The video frames are transformed to grayscale and normalized with respect to the global mean and variance. The training progress is similar to the word-level task, except using frame-level cross-entropy loss instead of sequence-level.

The audio-visual network is trained with 2s audio/video chunks using Adam~\cite{kingma2014adam} optimizer for 80 epochs with early stopping when there is no improvement on validation loss for 6 epochs. Initial learning rate is set to $1e^{-3}$ and halved during training if there is no improvement for 3 epochs on validation loss.


\subsection{Results and comparisons}
On both datasets, we first report the results of two typical oracle masks as well as two conventional audio based methods: uPIT-BLSTM and Conv-TasNet \cite{luo2018tasnet} on frequency and time-domain, respectively. The results of oracle mask are the upper-bound of the corresponding mask-based methods.  uPIT-BLSTM adopts a 3-layer Bi-LSTM structure and the PSM (phase sensitive mask \cite{erdogan2015phase}) is used as training target. For Conv-TasNet, we choose $K=40, S=20$ in audio encoder/decoder and a larger bottleneck size (384) in separation network in order to bring better performance. Results are shown in Table \ref{tab2}. It's worth to mention that the audio streams in LRS2 are not as clean as WSJ0, so the results reported on those audio-only models are worse than previous results on WSJ0-2mix, especially when the number of speakers increases.

\subsubsection{Results with different lip embeddings}
In Table \ref{tab3}, we evaluate the proposed audio-visual models with the word-level lip embeddings trained on the LRW dataset, which already leads to significant improvements compared with the audio-only methods in Table \ref{tab2}. As the resolutions of the video frames on LRS2 dataset do not match with LRW, we crop the center 70 $\times$ 70 pixel region of the images and then re-sample them to 112 $\times$ 112. Replacing word-level targets with CI-phones leads to better results, particularly on difficult tasks, i.e., three-speaker test set as shown in Table \ref{tab3}.

Normalization techniques are also critical in our experiments. In Table \ref{tab3}, replacing batch normalization (BN) with global layer normalization (gLN) leads to improvement on both two and three-speaker mixture datasets. In our preliminary experiments, the residual connection in the video block affects less on the final results. And by increasing the number of blocks in video encoders, no significant improvement is achieved, possibly due to the well trained visual features.

\begin{table}
\centering
\caption{Results of the model with different lip embeddings}
\label{tab3}
\begin{tabular}{ccccc}
\toprule
	\multirow{2}{*}{Dataset} & \multirow{2}{*}{\#Param} & \multirow{2}{*}{Embeddings} & \multicolumn{2}{c}{Si-SNR}       \\
	             &  &  &  BN & gLN      \\  \midrule
	\multirow{2}{*}{2spkr}      & \multirow{2}{*}{10.09M} & word  & 12.53    & 13.01   \\ 
	 &  & CI-phone & 12.76 &  \textbf{13.04}   \\  \hline
	\multirow{2}{*}{3spkr} & \multirow{2}{*}{10.09M} & word & 7.52 &  8.25  \\
	 &  & CI-phone  & 8.36 &  \textbf{9.41}   \\
\bottomrule
\end{tabular}
\end{table}

\begin{table}
\centering
\caption{Results of tuning number of fusion blocks}
\label{tab4}
\begin{tabular}{ccccc}
\toprule
	Dataset & $N_f$ & Embeddings  & Si-SNR \\ \midrule
	\multirow{2}{*}{2spkr} & 2 & \multirow{2}{*}{CI-phone} & 13.04 \\ 
	& 3 &  & \textbf{13.33} \\ \hline
	\multirow{3}{*}{3spkr} & 2 & \multirow{3}{*}{CI-phone} & 9.41 \\ 
	& 3 &  & \textbf{9.50} \\
	& 4 &  & 9.30 \\ \hline
	\multirow{3}{*}{3spkr} & 2 & \multirow{3}{*}{CD-phone} & 9.23 \\ 
	& 3 &  & \textbf{9.47} \\
	& 4 &  & 9.06 \\
\bottomrule
\end{tabular}
\end{table}

\subsubsection{Impact of separation networks}
The final results largely depend on the target speaker masks produced by the separation network. In addition to the normalization techniques mentioned in Section 3.3.1, we also tuned the number of $N_a$ and $N_f$ in the separation network. By fixing the total number of blocks (4 used in this work), increasing the value of $N_f$ brings more context information of fused features. Table \ref{tab4} illustrates that using $N_a = 1$ and $N_f = 3$ achieves the best performance.

Based on above discussion, we further compared CD/CI-phones on three-speaker dataset with different pair of $N_a$ and $N_f$. Results show that using CD-phone as training targets brings slightly worse result, although they are more suitable for acoustic modeling task. We therefore choose CI-phone in the following experiments.

\subsubsection{Comparision with the frequency-domain networks}
To further investigate the advances of the time-domain audio-visual approach, we trained a frequency-domain audio-visual separation model with a similar parameter size on the same dataset for comparison. We call it Conv-FavsNet in this paper. Conv-FavsNet removes the audio encoder from the proposed architecture and replaces the decoder with a linear layer, which transforms the output of convolutional blocks to TF-masks. We use linear spectrogram computed with 40ms hanning window and 10ms shift as input audio features and PSM as training target. The loss function for phase-sensitive spectrum approximation (PSA) is defined as:
\begin{equation} \label{}
    \mathcal{L} = \Vert \mathbf{s}_t \odot \max\{\cos(\angle \mathbf{s}_m - \angle \mathbf{s}_t), 0\} - \mathbf{s}_m \odot \mathbf{m}_t \Vert_2^2,
\end{equation}
where $\mathbf{m}_t$ denotes the estimated target speaker masks, and $\mathbf{s}_t, \mathbf{s}_m$ denote the magnitude of target speaker and mixture signal, respectively. Results are shown in Table \ref{tab6}. The noisy phase used in iSTFT certainly affects the quality of the reconstructed waveform as oracle phase can bring additional 3dB Si-SNR improvement which thus matches our time-domain audio-visual results.

\begin{table}
\centering
\caption{Results of frequency-domain audio-visual models}
\label{tab6}
\begin{tabular}{ccccc}
\toprule
Dataset  &Embeddings   & \#Param &Phase          & Si-SNR  \\ \midrule
\multirow{2}{*}{2spkr}  & \multirow{2}{*}{CI-phone}  & \multirow{2}{*}{10.03M}     &mix   & 10.36    \\
& & & oracle & 13.87 \\ \hline
\multirow{2}{*}{3spkr}  & \multirow{2}{*}{CI-phone}  & \multirow{2}{*}{10.03M}     &mix   & 6.16    \\
& & & oracle & 9.65 \\
\bottomrule
\end{tabular}
\end{table}

\subsubsection{Multi-speaker training}
For the biased/informed separation system, we find that models trained on hard tasks can work well on easy tasks. For example, in our experiments, the model trained on the three-speaker dataset also performs well on the two-speaker mixture signals. This motivates us to train models with a blend of two and three-speaker mixtures, which is called \emph{multi-speaker training} in \cite{isik2016single}. 

In fact, the architecture of the target isolating network is independent of the number of speakers. The label of training samples is determined by auxiliary features, e.g., lip embeddings in this work. This is applicable to the real-world scenarios because at most time the number of speakers is difficult to be detected. The performace of the models trained on two-speaker mixtures only, on three-speaker mixtures only, and using the multi-speaker training, are shown in Table \ref{tab5}. The model trained on three-speaker mixtures could generalize to two-speaker scenarios, but a little worse than the performance of the two-speaker model. The multi-speaker training brings improvement on two-speaker test set across all the setups in Table \ref{tab5}. The best configurations with multi-speaker training achieves 14.02dB and 9.92dB Si-SNR on two and three-speaker test sets, respectively.

\begin{table}
\centering
\caption{Results of models with multi-speaker training}
\label{tab5}
\begin{tabular}{ccccc}
\toprule
	\multirow{2}{*}{Training data} & \multirow{2}{*}{$N_f$} & \multirow{2}{*}{Norm} & \multicolumn{2}{c}{Si-SNR}       \\
	             &  &  &  2spkr & 3spkr      \\  \midrule
	2spkr & \multirow{3}{*}{2} & \multirow{3}{*}{BN} & 12.76 & -  \\
	3spkr &  &  & 11.52 & 8.36   \\  
    \{2,3\}spkr &  &  & \textbf{13.00} &  \textbf{8.38}   \\  \hline
	2spkr  & \multirow{3}{*}{2} & \multirow{3}{*}{gLN}  & 13.04 & -  \\
	3spkr &  &  & 12.74 & 9.41   \\  
    \{2,3\}spkr &  &  & \textbf{13.65} &  9.40   \\  \hline
	2spkr  & \multirow{3}{*}{3} & \multirow{3}{*}{gLN}  & 13.33 & -  \\
	3spkr &  &  & 12.86 & 9.50   \\  
    \{2,3\}spkr &  &  & \textbf{14.02} &  \textbf{9.92}   \\
\bottomrule
\end{tabular}
\end{table}

\section{Conclusions}

In this paper, we have proposed a time-domain audio-visual target speech separation architecture incorporating the raw waveform and the target speaker's video stream to predict the target speech waveform directly. Lip embedding extractor is pre-trained for extracting movement information from the video streams. We find that word-level and phoneme-level lip embeddings effectively benefit the separation network. Compared with the audio-only methods and the frequency-domain audio-visual method, the proposed approach improves more than 3dB and 4dB in terms of Si-SNR on two and three-speaker test sets, respectively. 


\bibliographystyle{IEEEbib}
\bibliography{strings,refs}

\end{document}